\documentclass[aps,psfig,epsfig,preprint]
{revtex4}
\usepackage{epsfig}
\usepackage{color}

\def\qq{Q\!\!\!\! Q}
\begin{document}
\draft
\title{
{\normalsize \hskip4.2in USTC-ICTS-05-06} \\{\bf Diquarks in
Nonaquark States
 }}
\author{Gui-Jun Ding\footnote{e-mail address: dinggj@mail.ustc.edu.cn},
Mu-Lin Yan\footnote{e-mail address: mlyan@ustc.edu.cn; corresponding
author.}}

\affiliation{}

\address{
 Interdisciplinary Center for Theoretical Study, University
of Science and Technology of China, \\ Hefei, Anhui 230026, China}

\begin{abstract}
  We study the nonaquark states $S^0(3115)$ and $S^+(3140)$ which are
reported by KEK-PS (Phys.Lett. B597 (2004) 236; nucl-ex/0310018) by
means of the quark model with diquark correlation. The nonaquark
states form
$\bf{1},\bf{8},\bf{10},\overline{\bf{10}},\bf{27},\overline{\bf{35}}$
$SU(3)$ multiplets. The flavor wave functions of all the nonaquark
states are constructed through the standard tensor technique. The
mass spectrum is studied by using Gell-Mann-Okubo mass formula. Some
nonaquark mass sum rules are obtained. We further investigate the
decay of $S^0(3115)$ and $S^+(3140)$ under the assumption of
"fall-apart" mechanism. It has been found that the main decay mode
is $\Sigma NN$ rather than $\Lambda NN$ which is consistent with
experiment. Also we have uniquely determine the flavor wave function
of $S^0(3115)$ which belong to $\bf{27}$-plet with the quantum
number $Y=2,I=1,I_z=-1$. Whereas the exotic states $S^+(3140)$ can
belong to either $\bf{27}$-plet or $\overline{\bf{35}}$-plet. In the
exact $SU(3)^{flavor}\times SU(3)^{color}\times SU(2)^{spin}$ limit,
both $S^0(3115)$ and $S^+(3140)$ belong to ${\bf 27}$-plet with
negative parity. We predict that its flavor structure can be
determined by measuring the branch fractions of its decay channels.
The experiments to check this prediction are expected.
\end{abstract}

\maketitle

\section{Introduction}

\noindent There is growing interest in exotic hadrons, which may
open new windows for understanding the hadronic structures and QCD
at low energies. Recently, the KEK-PS reported an tribaryon state
$S^0(3115)$\cite{kek1} in the reaction
\begin{equation}
\label{1.1}K^{-}+^{4}He\rightarrow S^0+p
\end{equation}
The mass of the state is $3117^{+1.5}_{-4.4}$MeV , the decay width
$\Gamma_{S^{0}} <$ 21 MeV, and the main decay mode is $\Sigma NN$
rather than $\Lambda NN$. The peak in the proton spectrum is over
the background with a significance level 13$\sigma$. A strange
tribaryon $S^+(3140)$ of charge +1 was also reported in the reaction
$K^{-}+^{4}He\rightarrow S^++n$ \cite{kek2}. The mass and decay
width of this exotic state are $M_{S^{+}}=3141\pm
3(stat.)^{+4}_{-1}(sys.)$ MeV and $\Gamma_{S^{+}}\leq 23$ MeV ,
which is about 25 MeV higher than $S^+(3115)$, and its significance
is 3.7$\sigma$. It also dominantly decay into $\Sigma NN$ rather
than $\Lambda NN$.
\par
The $S^0(3115)$ was first predicted by Akaishi and Yamazaki
\cite{first} as a deeply-bound kaonic state. Since the discovery of
$S^0(3115)$ and $S^+(3140)$, there has been some theoretical
discussion \cite{p1,p2,p3}, in Ref \cite{p1} these exotic states are
mainly analysized from MIT bag model, and they are identified as
kaonic bound state in Ref \cite{p2,p3}. Since the quark dynamics
could be regarded as a cornerstone for hadron physics, it is
interesting to investigate the nonaquark states by means of the
quark models. Various quark models have been used and proposed in
studying the pentaquark baryon state \cite{jaffe,lipkin,kim}. Here
we draw the spirit of Jaffe-- Wilczek's work\cite{jaffe}, since
there are evidences for strong diquark correlation in  the baryon
spectrum \cite{diquark}, and, especially, in  the light nonet-scalar
($J^{PC}=0^{++}$) meson spectrum. Their  masses are generally below
1000 MeV ($f_0(600),f_0(980),a_0(980),\kappa(800)$), and they do not
 favor the predictions of the q$\bar{\rm{q}}$-models, but
  favor the diquark-antidiquark's quite well. Diquark is a boson with color
$\overline{3}_c$, flavor $\overline{3}_{\rm{f}}$, and spin zero.
Diquark correlation is also the basis of color superconductivity in
dense quark matter which has not being observed experimentally. This
configuration is favored by one gluon exchange \cite{georgi,jaff2}
and by instanton interactions \cite{hooft,instanton}. It may play
important role in the exotic hadron physics. In this paper we try to
investigate nonaquark baryons by means of diquarks model, and to
learn what happens in the nonaquark case due to the strong diquark
correlation. Meanwhile in order to understand the decay of nonaquark
states, we suggest a decay mechanism which can qualitatively explain
the experiments and give us new predictions. This decay mechanism is
quite intuitive. To understand the structure of nonaquark, its mass
spectrum and the decay mechanism are the main aims of this paper.
\par
The paper is organized as follows, in the section II we study the
direct products of two diquarks states, four diquarks states , and
of four diquarks's plus one quark's. The irreducible tensors of the
allowed nonaquark states are derived. The flavor wave functions are
given by identifying the {$SU(3)$} tensors with the physical
tribaryon states. In Section III, the mass spectrum is derived by
using the Gell-Mann--Okubo mass formula. The Section IV devotes to
study the decays of $S^0(3115)$ and $S^+(3140)$ under the assumption
that the decays caused by a "fall-apart" mechanism. We find when
$S^0(3115)$ only belongs to a certain $\bf{27}$-plet, its main decay
mode is $\Sigma NN $ rather than $\Lambda NN$, and $S^+(3140)$ can
belongs to either $\bf{27}$-plet or $\overline{\bf{35}}$-plet. In
Section V, we briefly summary the results and give some discussions.

\section{The Flavor Wave Function of Nonaquark States }
\begin{figure}[hptb]
\begin{center}
\includegraphics*[width=8cm]{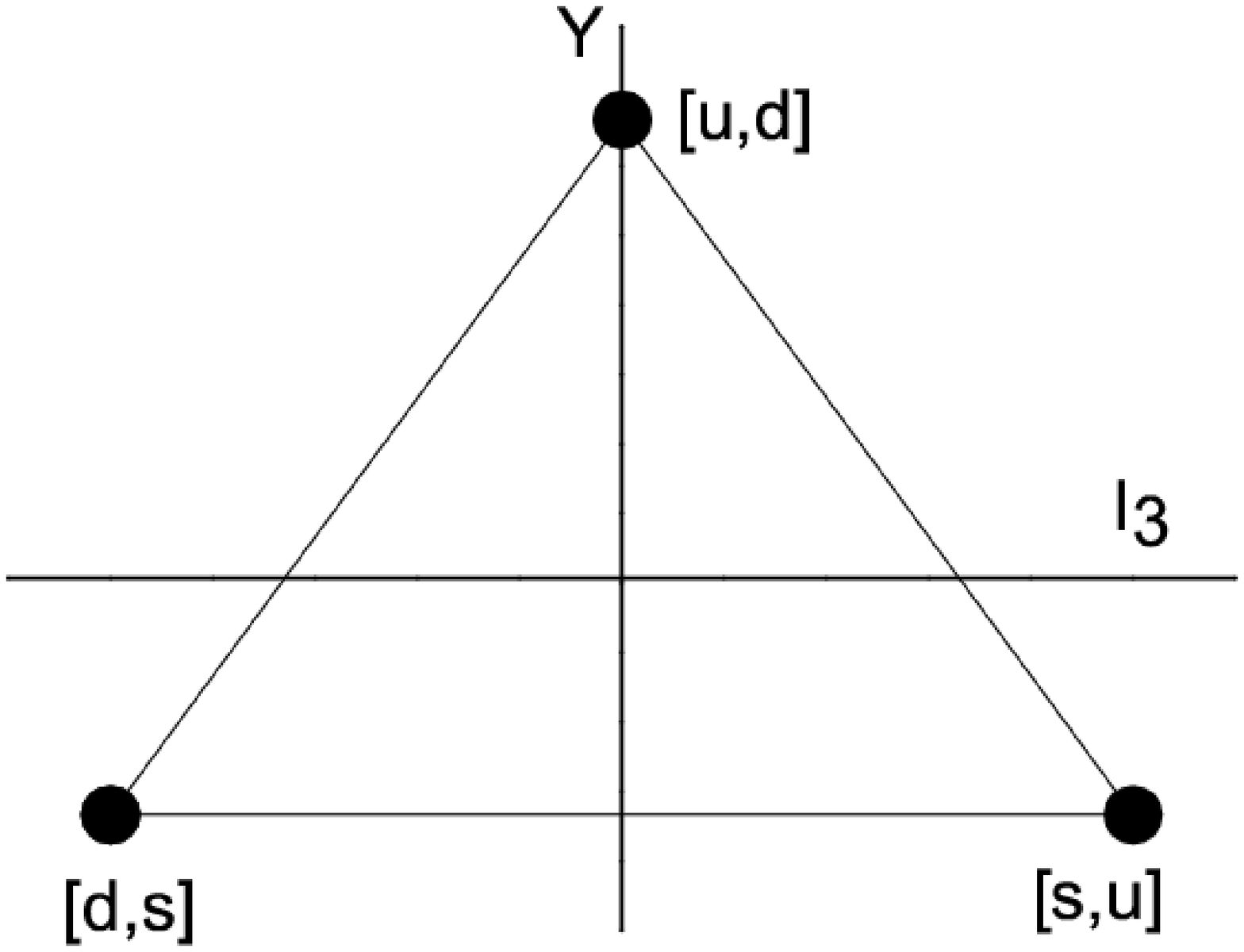}
\caption{$\overline{{\bf{3}}} $ diquark .}
\end{center}
\end{figure}
Since the diquark is in the  $\overline{3}_{\rm{f}}$, it has three
configurations in flavor space, which are shown in Fig1. We denote
them as
\begin{eqnarray}
\nonumber Q\!\!\!\! Q^{1}= \frac{1}{\sqrt{2}}[d,s]\\
\nonumber Q\!\!\!\! Q^{2}= \frac{1}{\sqrt{2}}[s,u]\\
\label{2.1} Q\!\!\!\! Q^{3}= \frac{1}{\sqrt{2}}[u,d]
\end{eqnarray}
where $u$, $d$, $s$ are respectively up quark,down quark and
strange quark. It is obvious that there are following
correspondences: $Q\!\!\!\! Q^{1}\leftrightarrow\overline{u},
Q\!\!\!\!Q^{2}\leftrightarrow\overline{d},
Q\!\!\!\!Q^{3}\leftrightarrow\overline{s}.$
\\
\begin{figure}
\begin{center}
\includegraphics*[width=8cm]{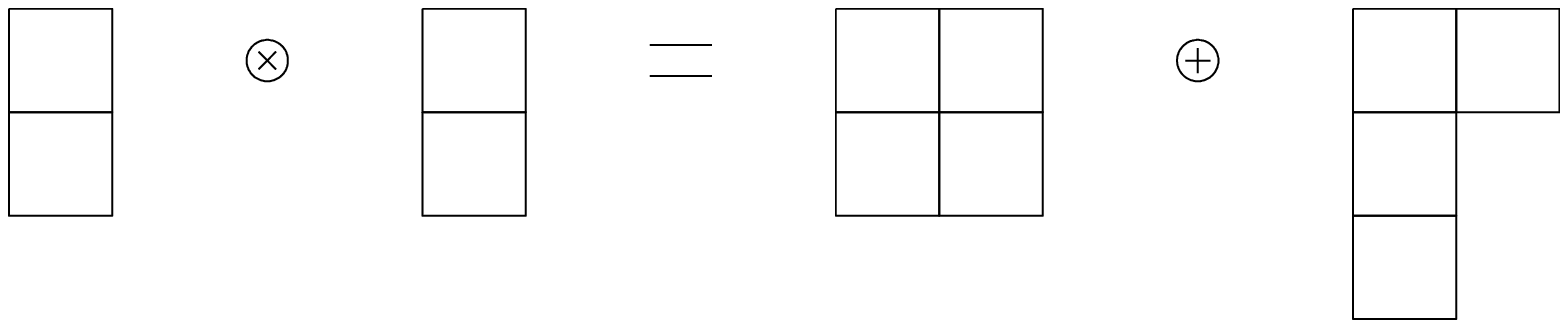}
\caption{direct product of two diquarks
$\overline{{\bf{3}}}\otimes\overline{{\bf{3}}}=\overline{\bf{6}}+\bf{3}
$ .}
\end{center}
\end{figure}

The tensors $T^{j_1,j_2,\cdot\cdot\cdot
j_q}_{i_1,i_2\cdot\cdot\cdot i_p}$ which are the bases for
irreducible representations of $SU(3)$ are totally symmetric to
both all q upper indices and all p low indices, and also are
traceless,
\begin{eqnarray}
\nonumber T^{j_1,j_2,\cdot\cdot\cdot j_q}_{i_1,i_2\cdot\cdot\cdot
i_p}&=&T^{j_2,j_1,\cdot\cdot\cdot j_q}_{i_1,i_2\cdot\cdot\cdot
i_p}=T^{j_1,j_2,\cdot\cdot\cdot j_q}_{i_2,i_1\cdot\cdot\cdot i_p}\\
\label{2.2}T^{i_1,j_2,\cdot\cdot\cdot j_q}_{i_1,i_2\cdot\cdot\cdot
i_p}&=&0.
\end{eqnarray}

Since
$\delta^{i}_{j},\hspace{2pt}\varepsilon^{ijk}\rm{and}\hspace{3pt}
\varepsilon_{ijk}$ are tensors, we can use them to raise, low or
contract indices when we construct new tensors that are bases of
irreducible representation from the direct product tensor. The
direct product of two diquarks is
\begin{equation}
\label{2.3}\qq^{i}\qq^{j}=\frac{1}{\sqrt{2}}S^{ij}+\frac{1}{2\sqrt{2}}\varepsilon^{ijk}T_k,
\end{equation}
with $S^{ij}=\frac{1}{\sqrt{2}}(\qq^{i}\qq^{j}+\qq^{j}\qq^{i})$,
$A^{ij}=\frac{1}{\sqrt{2}}(\qq^{i}\qq^{j}-\qq^{j}\qq^{i})$, and
 $T_k=\varepsilon_{ijk}A^{ij}$.
So the decomposition of the direct product of two diquarks is
$\overline{3}\otimes\overline{3}=\overline{6}\oplus 3$, which is
shown in Fig2. in the Young tabular.
\begin{figure}[hptb]
\begin{center}
\includegraphics*[23pt,310pt][554pt,500pt]{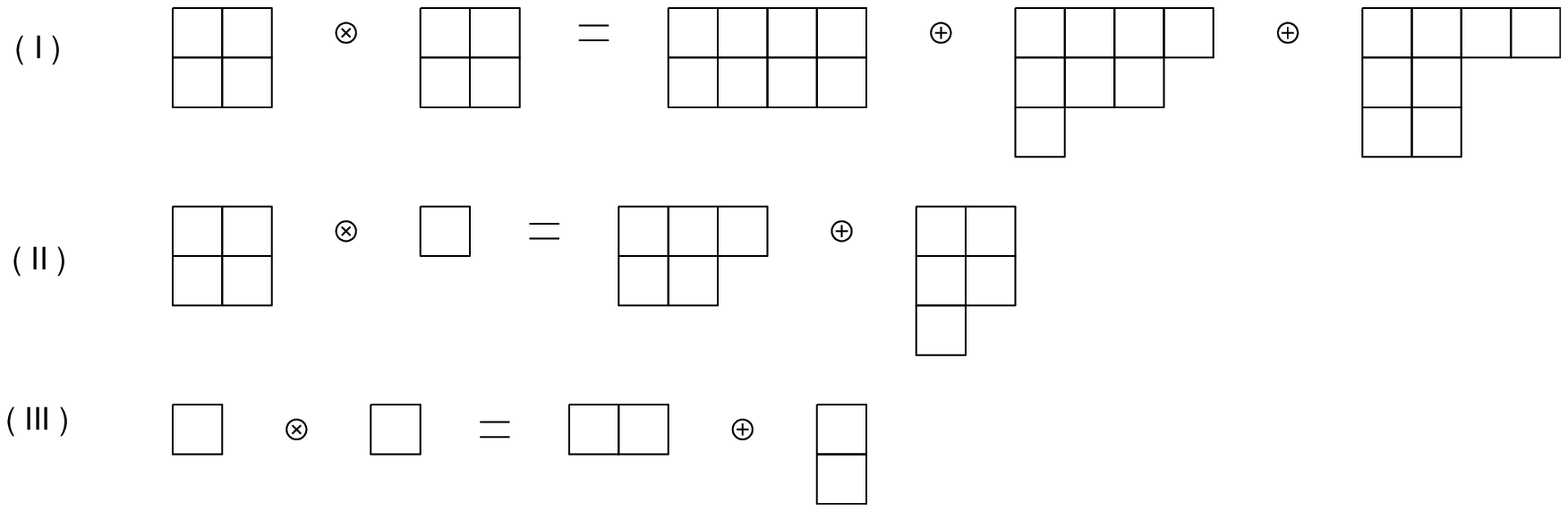}
\caption{The direct product of four diquarks: (I)
$\overline{{\bf{6}}}\otimes\overline{{\bf{6}}}=\overline{\bf{15}}_1+\overline{\bf{15}}_2+\bf{6}
$; (II)
$\overline{\bf{6}}\otimes\bf{3}=\overline{\bf{15}}_2+\overline{\bf{3}}$;
(III)$\bf{3}\otimes\bf{3}=\bf{6}\oplus\overline{\bf{3}}$.}
\end{center}
\end{figure}

Since the two diquarks can be decomposed into $\overline{\bf{6}}$
plus $\bf{3}$,  the direct product of four diquarks raises
$\overline{\bf{6}}\otimes \overline{\bf{6}}$,
$\overline{\bf{6}}\otimes \bf{3}$, $\bf{3}\otimes \overline{\bf{6}}$
and $\bf{3}\otimes \bf{3}$. And the corresponding Young tabular is
shown in Fig3. It is straightforward  that
\begin{eqnarray}
\nonumber &&(\qq^i\qq^j)(\qq^m\qq^n)\\
\nonumber&&=\frac{1}{2}S^{ij}S^{mn}+\frac{1}{4}(\varepsilon^{kij}T_{k}S^{mn}+\varepsilon^{kmn}S^{ij}T_{k})+\frac{1}{8}\varepsilon^{kij}\varepsilon^{lmn}T_kT_l\\
\nonumber&&=\frac{1}{2\sqrt{6}}T^{ijmn}+\frac{1}{4\sqrt{2}}(\varepsilon^{ajm}\delta^{n}_{b}\delta^{i}_{c}+\varepsilon^{ain}\delta^{m}_{b}\delta^{j}_{c})S^{bc}_{a}+\frac{1}{2\sqrt{6}}(\varepsilon^{aim}\varepsilon^{bjn}+\varepsilon^{ajm}\varepsilon^{bin})T_{ab}\\
\nonumber&&+\frac{1}{4}\varepsilon^{kij}[\tilde
T^{mn}_{k}+\frac{1}{\sqrt{2}}(\delta^{m}_{k}\delta^{n}_{a}+\delta^{n}_{k}\delta^{m}_{a})\tilde{Q}^{a}]+\frac{1}{4}\varepsilon^{kmn}[T^{ij}_{k}+\frac{1}{\sqrt{2}}(\delta^{i}_{k}\delta^{j}_{a}+\delta^{j}_{k}\delta^{i}_{a})Q^{a}]\\
\label{2.4}&&+\frac{1}{4}\varepsilon^{kij}\varepsilon^{lmn}(\sqrt{2}S_{kl}+\varepsilon_{kla}T^a)
\end{eqnarray}

where the tensors in the above formula are defined as followings.
\begin{equation}
\label{2.5}T^{i}=\frac{1}{8}\varepsilon^{ijk}(T_jT_k-T_kT_j)
\end{equation}
\begin{equation}
\label{2.6}Q^{i}=\frac{1}{\sqrt{8}}S^{ij}T_{j}
\end{equation}
\begin{equation}
\label{2.7}\tilde{Q}^{i}=\frac{1}{\sqrt{8}}T_jS^{ji}
\end{equation}
\begin{equation}
\label{2.8}S_{ij}=\frac{1}{4\sqrt{2}}(T_iT_j+T_jT_i)
\end{equation}
\begin{equation}
\label{2.9}T^{ijmn}=\frac{1}{\sqrt{6}}(S^{ij}S^{mn}+S^{mj}S^{in}+S^{in}S^{jm}+S^{mi}S^{jn}+S^{jn}S^{im}+S^{mn}S^{ij})
\end{equation}
\begin{equation}
\label{2.10}S^{jk}_{i}=\frac{1}{\sqrt{2}}\varepsilon_{imn}(S^{jm}S^{kn}+S^{km}S^{jn})
\end{equation}
\begin{equation}
\label{2.11}\tilde
T^{jk}_{i}=T_iS^{jk}-\frac{1}{\sqrt{2}}(\delta^{j}_{i}\delta^{k}_{m}+\delta^{k}_{i}\delta^{j}_{m})\tilde{Q}^{m}
\end{equation}
\begin{equation}
\label{2.12}T^{jk}_i=S^{jk}T_{i}-\frac{1}{\sqrt{2}}(\delta^{j}_{i}\delta^{k}_{m}+\delta^{k}_{i}\delta^{j}_{m})Q^m
\end{equation}
So the four diquarks product can be decomposed into
$\overline{{\bf{15}}}_1\oplus\overline{{\bf{15}}}_2(3)\oplus\overline{\bf{3}}(3)\oplus\bf{6}$,
where the numbers in the parentheses denote the degeneracy in each
multiplet. The tensors corresponding to $\overline{\bf{3}}$ are
$T^{i},Q^{i},\tilde Q^{i}$, the tensor $S_{ij}$ form the bases of
the irreducible representation $\bf{6}$, the tensor corresponding to
$\overline{\bf{15}}_1$ is $T^{ijmn}$, and the tensors
$S^{jk}_{i},\tilde T^{jk}_{i},T^{jk}_{i}$ are respectively the bases
of the irreducible representation $\overline{\bf{15}}_2$.
\vspace{1cm}
\begin{figure}[hptb]
\begin{center}
\includegraphics*[30pt,260pt][565pt,542pt]{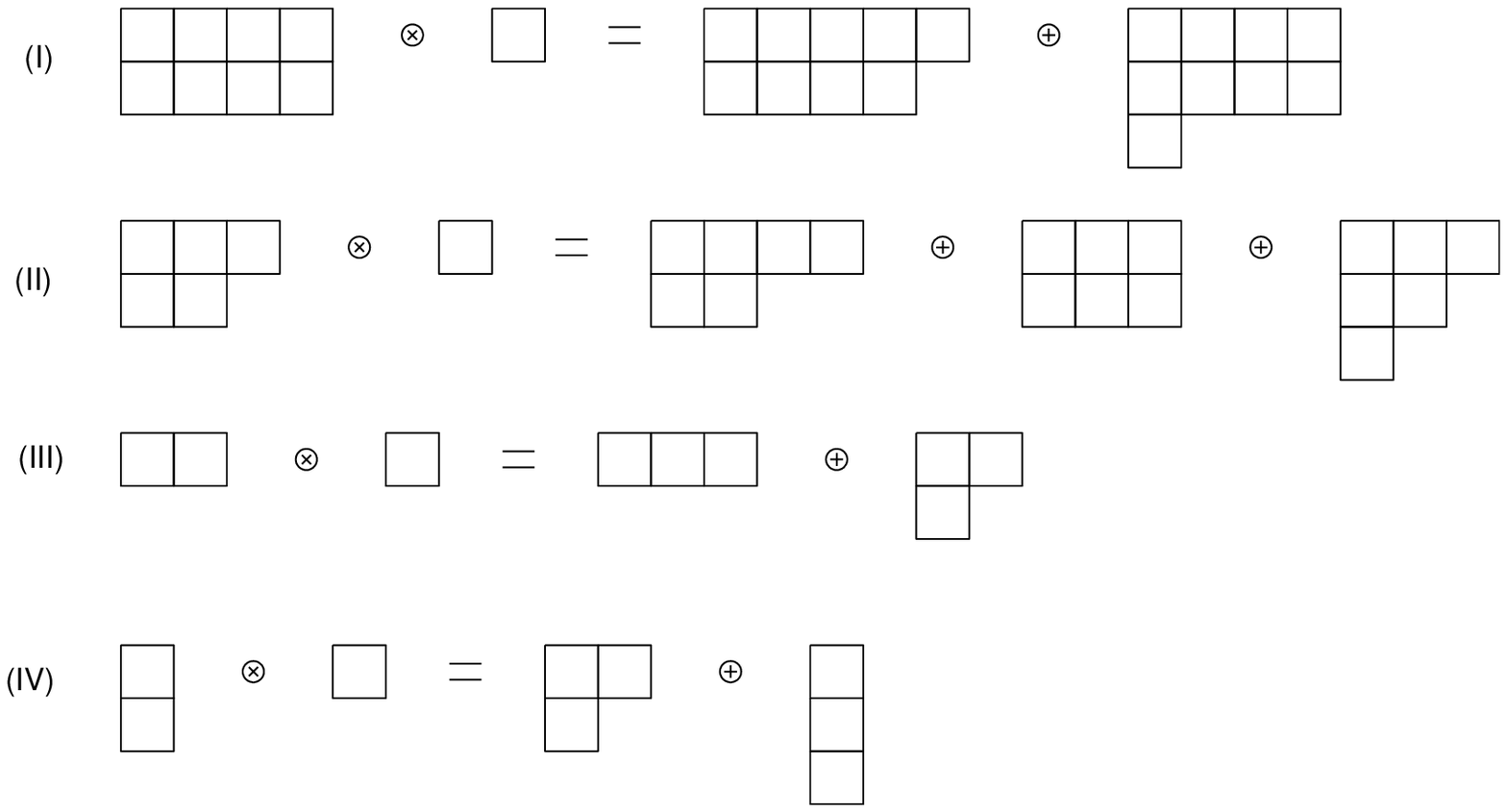}
\caption{direct product of four diquarks  and a quark (I)
$\overline{{\bf{15}}}_1\otimes\bf{3}=\overline{\bf{35}}+\overline{\bf{10}}
$; (II)
$\overline{\bf{15}}_2\otimes\bf{3}=\bf{27}+\overline{\bf{10}}+\bf{8}$;
~(III)
$\bf{6}\otimes\bf{3}=\bf{10}\oplus\bf{8}$;~(IV)$\overline{\bf{3}}\otimes\bf{3}=\bf{8}\oplus\bf{1}$.}
\end{center}
\end{figure}

\begin{center}
{\subsection{Nonaquark states}}
\end{center}
Since quark is in the fundamental representation $\bf{3}$, when the
four diquarks form the irreducible representation
$\overline{\bf{3}}$,  the nonaquark state must be in the
representation $\overline{\bf{3}}\otimes\bf{3}=\bf{8}+\bf{1}$. This
means the nonaquark state can either in the octet or in the singlet.
We use ${\cal T}^{i}$ to stand for $T^i,Q^i,\tilde Q^{i}$, then the
tensor product ${\cal T}^i q_n$ can be  decomposed as follows
\begin{equation}
\label{2.13} {\cal T}^i
q_n=\sqrt{2}(P^{i}_{n}+\frac{1}{\sqrt{3}}\delta^{i}_{n}S),
\end{equation}
where $S=\frac{1}{\sqrt{6}}{\cal T}^mq_m$,
$P^{i}_{n}=\frac{1}{\sqrt{2}}({\cal
T}^iq_n-\sqrt{\frac{2}{3}}\delta^{i}_{n}S)$.  $P^{i}_{n}$ stand for
the nonaquark octet, and that $S$ stands for the nonaquark singlet.

When the four diquarks are in the representation
$\overline{\bf{6}}\otimes\overline{\bf{6}}~(=\overline{{\bf
15}}_1\bigoplus \overline{{\bf 15}}_2\bigoplus {\bf 6}$ (see Fig.
3)),~ they can form the irreducible representative of $\bf{6}$.
 Since $\bf{6}\otimes\bf{3}=\bf{10}\oplus\bf{8}$, the nonaquark states
 can be in decuplet or octet.
\begin{equation}
\label{2.14}S_{ij}q_n=\frac{1}{\sqrt{3}}[T_{ijk}+\varepsilon_{mjn}P^{m}_{i}+\varepsilon_{min}P^{m}_{j}]
\end{equation}
with $T_{ijk}=\frac{1}{\sqrt{3}}[S_{ij}q_n+S_{in}q_j+S_{jn}q_i]$,
$P^{j}_{i}=\frac{1}{\sqrt{3}}\varepsilon^{jab}S_{ia}q_b$ and
$T_{ijk}, P^{j}_{i}$ respectively correspond to the nonaquark
decuplet and octet.

Again, for the four-diquarks states in
$\overline{\bf{6}}\otimes\overline{\bf{6}}$, they can also form
$\overline{\bf{15}}_1$. The direct product
$\overline{\bf{15}}_1\otimes\bf{3}$ can be reduced as follows
\begin{equation}
\label{2.15}T^{ijkl}q_n=T^{ijkl}_{n}+\frac{1}{\sqrt{6}}(\delta^{i}_{n}\delta^{j}_{b}\delta^{k}_{c}\delta^{l}_{d}+\delta^{j}_{n}\delta^{i}_{b}\delta^{k}_{c}\delta^{l}_{d}+\delta^{k}_{n}\delta^{i}_{b}\delta^{j}_{c}\delta^{l}_{d}+\delta^{l}_{n}\delta^{i}_{b}\delta^{j}_{c}\delta^{k}_{d})D^{bcd}
\end{equation}
where
$T^{ijkl}_{n}=T^{ijkl}q_n-\frac{1}{\sqrt{6}}(\delta^{i}_{n}D^{jkl}+\delta^{j}_{n}D^{ikl}+\delta^{l}_{n}D^{ijk}),
\;D^{ijk}=\frac{1}{\sqrt{6}}T^{ijkn}q_n$, $T^{ijkl}_{n}$ and
$D^{ijk}$ respectively mean that the nonaquark state belong to
$\overline{\bf{35}}-$plet and $\overline{\bf{10}}$-plet.

Finally, we consider the case of that the four-diquarks states form
$\overline{\bf{15}}_2$. The $\overline{\bf{15}}_2-$ four-diquarks
states have three irreducible representatives: they can be in the
product of
$\overline{\bf{6}}\otimes\bf{3}$,~~$\bf{3}\otimes\overline{\bf{6}}$,~
or $\overline{\bf{6}}\otimes\overline{\bf{6}}$. The corresponding
tensors are $S^{jk}_{i}$,$\tilde T^{jk}_{i}$ or $T^{jk}_{i}$
respectively. Using ${\cal T}^{jk}_{i}$ to denote each one of them,
then we have
\begin{equation}
\label{2.16}{\cal
T}^{jk}_{i}q_n=\sqrt{2}T^{jk}_{in}+\sqrt{\frac{2}{3}}\varepsilon_{inm}D^{mjk}
+\frac{4}{\sqrt{15}}(\delta^{k}_{n}\delta^{j}_{m}+\delta^{k}_{m}\delta^{j}_{n})P^{m}_{i}
-\frac{1}{\sqrt{15}}(\delta^{k}_{i}\delta^{j}_{m}+\delta^{k}_{m}\delta^{j}_{i})P^{m}_{n},
\end{equation}
where
\begin{eqnarray}
\nonumber P^{i}_{j}&=&\frac{1}{\sqrt{15}} {\cal T}^{ik}_{j}q_k\\
\nonumber
D^{ijk}&=&\frac{1}{24}(\varepsilon^{jab}{\cal T}^{km}_{a}q_b+\varepsilon^{kab}{\cal T}^{jm}_{a}q_b+\varepsilon^{mab}{\cal T}^{jk}_{a}q_b)\\
\label{2.17}T^{jk}_{in}&=&\frac{1}{2\sqrt{2}}({\cal
T}^{jk}_{i}q_n+{\cal
T}^{jk}_{n}q_i)-\frac{\sqrt{30}}{20}(\delta^{k}_{i}P^{j}_{n}+\delta^{k}_{n}P^{j}_{i}
+\delta^{j}_{i}P^{k}_{n}+\delta^{j}_{n}P^{k}_{i}).
\end{eqnarray}
$T^{jk}_{in}$ represent the nonaquark $\overline{35}$-plet, and
$D^{ijk}, P^{i}_{j}$ correspond to nonaquark $\overline{10}$-plet
and octet. The Young tabular of the tensor decomposit
ion are shown
in Fig4.

Now we take the color symmetry $SU(3)^{color}$ into the account.
When the $SU(3)^{flavor}\times SU(3)^{color}\times SU(2)^{spin}$
serves as an exact symmetry, then due to the boson statistics the
full combined ({\it flavor $\times$ color $\times$ spin $\times$
space}) wave-functions of four diquarks states must be symmetric.
Since the spin of the diquark is zero, the {\it spin}-wave-function
is trivially to be symmetric. Thus, the ({\it space $\times$ flavor
$\times$ color}) wave function must be symmetric. Under this
constraint, only two choices are available:

1)The {\it space}-wave-functions are symmetric: In this case, the
relative angular momentum $\ell$ is even and starts from $\ell=0$,
and then the ({\it flavor $\times$ color}) wave-functions are
symmetric. They must therefore belong to the $\overline{495}$
dimensional irreducible representation of SU(9)$\supset$
SU(3)$^{flavor }$$\times$ SU(3)$^{color}$. The reduction of the
SU(9) irreducible representation  $\overline{495}$ with respect to
SU(3)$^{flavor }$$\times$ SU(3)$^{color}$ \cite{group} is
\begin{equation}
\label{2.18}\overline{495}=(\overline{15}_1,\overline{15}_1)+(\overline{15}_2,\overline{15}_2)+(\bar{3},\bar{3})+(6,6).
\end{equation}
We can see only the four diquarks's  states
$(flavor,color)=(\bar{3},\bar{3})$ can combine with the ninth quark
with $(flavor,color)=(3,3)$ to form a color singlet. Because
$\bar{3}\otimes 3=8\oplus1$, there may exist nonaquark nonet with
positive parity. But the states $S^{0}(3115)$ and $S^{+}(3140)$ can
not belong to this nonet, because the hypercharge of these two
states is $Y=2$ , while the maximum of hypercharge of the nonet is 1
.

2) The {\it space}-wave-functions are antisymmetric: This means that
the relative angular momentum $\ell$ is odd and starts from
$\ell=1$, and then the ({\it flavor $\times$ color}) wave-functions
are antisymmetric. And they must belong to  the 126 dimensional
irreducible representation of SU(9)$\supset$ SU(3)$^{flavor
}$$\times$ SU(3)$^{color}$. The reduction of the SU(9) irreducible
representation 126 with respect to SU(3)$^{flavor }$$\times$
SU(3)$^{color}$ is
\begin{equation}
\label{2.19}126=(6,6)+(\overline{15}_2,\bar{3})+(\bar{3},\overline{15}_2)
\end{equation}
So only when the four diquarks in the state
$(flavor,color)=(\overline{15}_2,\bar{3})$, they can combine with
the ninth quark to form a color singlet hadron. Because
$\overline{15}_2\otimes 3=27\oplus \overline{10}\oplus 8$, the
nonaquark can only in flavor multiplet 27-plet,
$\overline{10}$-plet, 8-plet, and the nonaquark state can not be
flavor 35-plet. This is a rigorous result  when the flavor symmetry
is exactly. Considering, however, that $SU(3)^{flavor}$ is an
approximative symmetry which will lead to $SU(3)^{flavor }\times
SU(3)^{color}\times SU(2)^{spin}$ to be approximate, so we can not
completely rule out 35-plet.
\par
 From the quantum number of $S^{0}(3115)$ and
$S^{+}(3140)$, they possibly belong to the $\bf{27}$-plet or
$\overline{\bf{35}}$-plet. In the exact $SU(3)^{flavor}\times
SU(3)^{color}\times SU(2)^{spin}$ limit,
 both $S^{0}(3115)$ and $S^{+}(3140)$ belong to {\bf 27}-plet whose lowest angular momentum is
$\ell=1$ (they are  P-wave states), and the corresponding weight
diagram is Fig6. The Fig7 is the weight diagram for 35-plet. In the
Figures, we show the names of these exotic states, with the
subscripts that are the representation-dimensions and the isospin of
the particle. Their superscript is the charge of the state.
\par
In tensor representations, the number of lower indices of $T_{i_1,
\dots, i_p}^{j_1, \dots,k_q}$ is $p$ and that of upper indices is
$q$. Now we suppose that among its lower indices the numbers of 1,
2, and 3 are $p_1$, $p_2$, and $p_3$, respectively, and that among
upper indices it has $q_1$ 1, $q_2$ 2, and $q_3$ 3. Then we have
$p_1+p_2+p_3 = p$ and $q_1+q_2+q_3=q$. The irreducible tensor is an
eigenstate of hypercharge $Y$ and the third component of isospin
$I_3$ with the eigenvalues \cite{lowb,closeb,inpp}
\begin{eqnarray}
\nonumber Y &=& p_1 - q_1 + p_2 - q_2 - \frac23 (p-q)  \\
\label{2.20} I_3 &=& \frac12 (p_1-q_1) - \frac12 (p_2-q_2).
\end{eqnarray}
The charge of the particle is obtained from the Gell-Mann--Nishijima
formula, $Q = I_3 + Y/2$. By this way, we can match the SU(3)
tensors to the physical baryon states. \vskip0.3in

\begin{center}
{\subsection{The wavefunction of Nonaquark
$\overline{\bf{27}}$-plet}}
\end{center}\vskip-0.5in
It is straightforward to write out the wave functions of nonaquark
$\overline{\bf{27}}$-plets in the flavor space by means of the
irreducible  representation tensors. The $S^{+}_{27,1}$ and
$S^{0}_{27,1}$ read
\begin{equation}
\nonumber
\label{2.29}S^{+}_{27,1}=\frac{1}{\sqrt{2}}T^{33}_{12}=\frac{1}{4}(\mathcal{T}^{33}_{1}q_2+\mathcal{T}^{33}_2q_1)
\end{equation}
\begin{equation}
\label{2.30}S^{0}_{27,1}=\frac{1}{2}T^{33}_{22}=\frac{1}{2\sqrt{2}}\mathcal{T}^{33}_{2}q_2,
\end{equation}
where $\mathcal{T}^{ij}_k$ stands for the tensors of four-diquarks
states. We now provide the explicit expressions of $S^{+}_{27,1}$
and $S^{0}_{27,1}$ for the each irreducible representatives of
$\overline{\bf{15}}_2$ in order:

\begin{enumerate}

\item The  case of
$\overline{\bf{15}}_2\subset\overline{\bf{6}}\otimes\overline{\bf{6}}$:
In this case ${\cal T}^{jk}_{i}$ is $S^{jk}_{i}$ which has been
defined in Eq.(\ref{2.10}). So,
\begin{eqnarray}
\nonumber&& S^{33}_{1}=\frac{1}{2\sqrt{2}}([u,d][s,u][u,d][u,d]+[s,u][u,d][u,d][u,d]-[u,d][u,d][u,d][s,u]\\
\label{2.31}&&-[u,d][u,d][s,u][u,d]),
\end{eqnarray}
\begin{eqnarray}
\nonumber
&&S^{33}_{2}=\frac{1}{2\sqrt{2}}([u,d][u,d][u,d][d,s]+[u,d][u,d][d,s][u,d]-[u,d][d,s][u,d][u,d]\\
\label{2.32}&&-[d,s][u,d][u,d][u,d]),
\end{eqnarray}
the wave function of $S^{+}_{27_1,1}$ and $S^{0}_{27_1,1}$ are
\begin{eqnarray}
\nonumber S^{+}_{27_1,1}&=&\frac{1}{4}(S^{33}_{1}q_2+S^{33}_{2}q_1)\\
\nonumber&=&\frac{1}{8\sqrt{2}}([u,d][s,u][u,d][u,d]d+[s,u][u,d][u,d][u,d]d-[u,d][u,d][u,d][s,u]d\\
\nonumber&&-[u,d][u,d][s,u][u,d]d+[u,d][u,d][u,d][d,s]u+[u,d][u,d][d,s][u,d]u\\
\label{2.33}&&-[u,d][d,s][u,d][u,d]u-[d,s][u,d][u,d][u,d]u);
\end{eqnarray}
\begin{eqnarray}
\nonumber S^{0}_{27_1,1}&=&\frac{1}{2\sqrt{2}}S^{33}_{2}q_2\\
\nonumber&=&\frac{1}{8}([u,d][u,d][u,d][d,s]d+[u,d][u,d][d,s][u,d]d-[u,d][d,s][u,d][u,d]d\\
\label{2.34}&&-[d,s][u,d][u,d][u,d]d).
\end{eqnarray}

\item The  case of
$\overline{\bf{15}}_2\subset\bf{3}\otimes\overline{\bf{6}}$: The
${\cal T}^{jk}_{i}$ is $\tilde{T}^{jk}_{i}$ which is defined in
Eq.(\ref{2.11}), and
\begin{equation}
\label{2.35}\tilde{T}^{33}_{1}=T_1S^{33}=\frac{1}{2}([s,u][u,d][u,d][u,d]-[u,d][s,u][u,d][u,d])
\end{equation}
\begin{equation}
\label{2.36}\tilde{T}^{33}_{2}=T_2S^{33}=\frac{1}{2}([u,d][d,s][u,d][u,d]-[d,s][u,d][u,d][u,d])
\end{equation}
then, the wave functions of $S^{+}_{27_2,1}$ and $S^{0}_{27_2,1}$
are
\begin{eqnarray}
\nonumber S^{+}_{27_2,1}&=&\frac{1}{4}(\tilde T^{33}_{1}q_2+\tilde
T^{33}_{2}q_1)\\
\nonumber
&=&\frac{1}{8}([s,u][u,d][u,d][u,d]d-[u,d][s,u][u,d][u,d]d+[u,d][d,s][u,d][u,d]u\\
\label{2.37}&&-[d,s][u,d][u,d][u,d]u);
\end{eqnarray}
\begin{equation}
\label{2.38} S^{0}_{27_2,1}=\frac{1}{2\sqrt{2}}\tilde
T^{33}_{2}q_2=\frac{1}{4\sqrt{2}}([u,d][d,s][u,d][u,d]d-[d,s][u,d][u,d][u,d]d).
\end{equation}

\item The case of
$\overline{\bf{15}}_2\subset\overline{\bf{6}}\otimes\bf{3}$: In this
case ${\cal T}^{jk}_{i}$ is the tensor $T^{jk}_{i}$ defined in
Eq.(\ref{2.12}), obviously
\begin{equation}
\label{2.39}T^{33}_{1}=S^{33}T_1=\frac{1}{2}([u,d][u,d][s,u][u,d]-[u,d][u,d][u,d][s,u])
\end{equation}
\begin{equation}
\label{2.40}T^{33}_{2}=S^{33}T_{2}=\frac{1}{2}([u,d][u,d][u,d][d,s]-[u,d][u,d][d,s][u,d])
\end{equation}
The wave functions of the two states are
\begin{eqnarray}
\nonumber S^{+}_{27_3,1}&=&\frac{1}{4}(T^{33}_{1}q_2+T^{33}_{2}q_1)\\
\nonumber&=&\frac{1}{8}([u,d][u,d][s,u][u,d]d-[u,d][u,d][u,d][s,u]d+[u,d][u,d][u,d][d,s]u\\
\label{2.41}&&-[u,d][u,d][d,s][u,d]u)
\end{eqnarray}
\begin{eqnarray}
\nonumber S^{0}_{27_3,1}&=&\frac{1}{2\sqrt{2}}T^{33}_{2}q_2\\
\label{2.42}
&=&\frac{1}{4\sqrt{2}}([u,d][u,d][u,d][d,s]d-[u,d][u,d][d,s][u,d]d).
\end{eqnarray}

\end{enumerate}

 \vspace{0.5cm}
\begin{center}
{\subsection{The wave function of Nonaquark
$\overline{\bf{35}}$-plet}}
\end{center}
Since $\overline{{\bf 35}}$ can not be completely excluded (see the
subsection II.A) , we should also discuss it's wavefunction for
completeness. It is easy to identify
\begin{eqnarray}
\nonumber&& T^{1333}_{2}=-\sqrt{6}S^{0}_{\overline{35},1},~~~~
T^{1333}_{1}=-\sqrt{3}S^{+}_{\overline{35},1}-\sqrt{2}S^{+}_{\overline{35}}\\
\label{2.21}&&T^{2333}_{2}=\sqrt{3}S^{+}_{\overline{35},1}-\sqrt{2}S^{+}_{\overline{35}},~~T^{3333}_{3}=2\sqrt{2}S^{+}_{\overline{35}},~~T^{2333}_{1}=-\sqrt{6}S^{++}_{\overline{35},1}
\end{eqnarray}
then
\begin{equation}
\label{2.22} \left\{
 \begin{array}{l}
 S^{0}_{\overline{35},1}=-\frac{1}{\sqrt{6}}T^{1333}_{2}\\
  S^{+}_{\overline{35},1}=\frac{1}{2\sqrt{3}}(T^{2333}_{2}-T^{1333}_{1})\\
  S^{+}_{\overline{35}}=\frac{1}{2\sqrt{2}}T^{3333}_{3}
 \end{array}
 \right.
\end{equation}
and $T^{ijkl}_{n}$  is defined in Eq.(\ref{2.15}). It is easy to see
\begin{eqnarray}
\nonumber
T^{1333}&=&\frac{1}{\sqrt{6}}(S^{13}S^{23}+S^{33}S^{13}+S^{13}S^{33}+S^{31}S^{33}+S^{33}S^{13}+S^{33}S^{13})\\
\nonumber&=&\frac{3}{\sqrt{6}}(S^{13}S^{33}+S^{33}S^{13})\\
\nonumber&=&\frac{3}{4\sqrt{6}}([d,s][u,d][u,d][u,d]+[u,d][d,s][u,d][u,d]\\
\label{2.23}&&+[u,d][u,d][d,s][u,d]+[u,d][u,d][u,d][d,s])
\end{eqnarray}
\begin{eqnarray}
\nonumber
T^{2333}&=&\frac{3}{\sqrt{6}}(S^{23}S^{33}+S^{33}S^{23})\\
\nonumber&=&\frac{3}{4\sqrt{6}}([s,u][u,d][u,d][u,d]+[u,d][s,u][u,d][u,d]\\
\label{2.24}&&+[u,d][u,d][s,u][u,d]+[u,d][u,d][u,d][s,u])
\end{eqnarray}
\begin{equation}
\label{2.25}T^{3333}=\sqrt{6}S^{33}S^{33}=\frac{\sqrt{6}}{2}[u,d][u,d][u,d][u,d]
\end{equation}
and the wave function of
$S^{0}_{\overline{35},1},S^{+}_{\overline{35},1},S^{+}_{\overline{35}}$
are as followings
\begin{eqnarray}
\nonumber
S^{0}_{\overline{35},1}&=&-\frac{1}{\sqrt{6}}T^{1333}_{2}=-\frac{1}{\sqrt{6}}T^{1333}q_{2}\\
\nonumber&=&-\frac{1}{8}([d,s][u,d][u,d][u,d]+[u,d][d,s][u,d][u,d]\\
\label{2.26}&&+[u,d][u,d][d,s][u,d]+[u,d][u,d][u,d][d,s])d
\end{eqnarray}

\begin{eqnarray}
\nonumber S^{+}_{\overline{35},1}&=&\frac{1}{2\sqrt{3}}(T^{2333}_{2}-T^{1333}_{1})=\frac{1}{2\sqrt{3}}(-T^{1333}q_1+T^{2333}q_2)\\
\nonumber
&=&\frac{1}{8\sqrt{2}}(-[d,s][u,d][u,d][u,d]u-[u,d][d,s][u,d][u,d]u-[u,d][u,d][d,s][u,d]u\\
\nonumber &&-[u,d][u,d][u,d][d,s]u+[s,u][u,d][u,d][u,d]d+[u,d][s,u][u,d][u,d]d\\
\label{2.27}&&+[u,d][u,d][s,u][u,d]d+[u,d][u,d][u,d][s,u]d)
\end{eqnarray}

\begin{eqnarray}
\nonumber S^{+}_{\overline{35}}&=&\frac{1}{2\sqrt{2}}T^{3333}_{3}=\frac{1}{6\sqrt{2}}(-2T^{1333}q_1-2T^{2333}q_2+T^{3333}q_3)\\
\nonumber&=&-\frac{1}{8\sqrt{3}}([d,s][u,d][u,d][u,d]u+[u,d][d,s][u,d][u,d]u+[u,d][u,d][d,s][u,d]u\\
\nonumber&&+[u,d][u,d][u,d][d,s]u+[s,u][u,d][u,d][u,d]+[u,d][s,u][u,d][u,d]\\
\label{2.28}&&+[u,d][u,d][s,u][u,d]+[u,d][u,d][u,d][s,u]-2[u,d][u,d][u,d][u,d]s)
\end{eqnarray}
\vspace{0.2cm}

\begin{center}
{\section{The Mass Spectrum of Nonaquark States}}
\end{center}
Since all the particles belonging to an irreducible representation
of $SU(3)$ are degenerate in $SU(3)$ symmetry limit, it is
necessary to introduce the $SU(3)$ symmetry breaking terms into
the Hamiltonian in order to obtain the mass splitting. The
Hamiltonian that breaks $SU(3)$ symmetry but still preserves the
isospin symmetry and hypercharge is proportional to the Gell-Mann
matrix $\lambda_8$, and the baryon mass can be obtained by
constructing $SU(3)$ singlet term including the hypercharge
tensor, in this way we obtain the Gell-Mann-Okubo mass formula:
\begin{equation}
\label{3.1}M=M_0+\alpha Y+\beta D^{3}_{3}
\end{equation}
where $M_0$ is a common mass of a given multiplet and
$D^{3}_{3}=I(I+1)-\frac{Y^2}{4}-\frac{C}{6}$ with
$C=2(p+q)+\frac{2}{3}(p^2+pq+q^2)$ for the (p,q) representation.
$\alpha$ and $\beta$ are mass constant that are in principle
different for different multiplets. Using these constants, we can
obtain the masses of all the baryons within the multiplet. Note that
in this picture the isospin is conserved.
\\
\begin{center}
{\subsection{The mass spectrum of Nonaquark $\bf{27}$-plet}}
\end{center}
In the case of $\bf{27}$-plet, $p=q=2$ and the corresponding weight
diagram is Fig.(6). By using the Gell-Mann-Okubo mass formula
Eq.(\ref{3.1}) we can get all the masses of these states
\begin{eqnarray}
\nonumber
&&M_{S_{27,1}}=M_{27}+2\alpha_{27}-\frac{5}{3}\beta_{27}\;,~~M_{N_{27,\frac{3}{2}}}=M_{27}+\alpha_{27}+\frac{5}{6}\beta_{27}\;,~~M_{N_{27,\frac{1}{2}}}=M_{27}+\alpha_{27}-\frac{13}{6}\beta_{27}\\
\nonumber&&M_{\Sigma_{27,2}}=M_{27}+\frac{10}{3}\beta_{27}\;,~~M_{\Sigma_{27,1}}=M_{27}-\frac{2}{3}\beta_{27}\;,~~M_{\Lambda_{27}}=M_{27}-\frac{8}{3}\beta_{27}\\
\nonumber&&M_{\Xi_{27,\frac{3}{2}}}=M_{27}-\alpha_{27}+\frac{5}{6}\beta_{27}\;,~~M_{\Xi_{27,\frac{1}{2}}}=M_{27}-\alpha_{27}-\frac{13}{6}\beta_{27}\\
\label{3.6}&&M_{\Omega_{27,1}}=M_{27}-2\alpha_{27}-\frac{5}{3}\beta_{27}
\end{eqnarray}
The Gell-Mann-Okubo mass relation for 27--plet nonaquark baryons is
\begin{equation}
\label{3.7}3M_{\Lambda_{27}}+M_{\Sigma_{27,1}}=2(M_{
N_{27,\frac{1}{2}}}+M_{\Xi_{27,\frac{1}{2}}})=4M_{27}-\frac{26}{3}\beta_{27}
\end{equation}
The equal mass space relations also exist in the two separate
sectors: ($\Omega_{27,1},\Xi_{27,\frac{3}{2}},\Sigma_{27,2}$) and
($\Sigma_{27,2},N_{27,\frac{3}{2}},S_{27,1}$). Their mass relations
are as follows
\begin{equation}
\label{3.8}M_{\Omega_{27,1}}-M_{\Xi{27,\frac{3}{2}}}=M_{\Xi{27,\frac{3}{2}}}-M_{\Sigma_{27,2}}=-\alpha_{27}-\frac{5}{2}\beta_{27}
\end{equation}
\begin{equation}
\label{3.9}M_{\Sigma_{27,2}}-M_{N_{27,\frac{3}{2}}}=M_{N_{27,\frac{3}{2}}}-M_{S_{27,1}}=-\alpha_{27}+\frac{5}{2}\beta_{27}
\end{equation}
From Eq.(\ref{3.6}) we can further obtain some relation between mass
of these states
\begin{eqnarray}
\nonumber&&M_{\Lambda_{27}}+M_{S_{27,1}}=2M_{N_{27,\frac{1}{2}}}\;,~~M_{\Sigma_{27,2}}+M_{S_{27,1}}=2M_{N_{27,\frac{3}{2}}}\;,~3M_{\Sigma_{27,1}}+3M_{S_{27,1}}=4M_{N_{27,\frac{1}{2}}}+2M_{N_{27,\frac{3}{2}}}\\
\nonumber&&3M_{\Xi_{27,\frac{3}{2}}}+6M_{S_{27,1}}=5M_{N_{27,\frac{1}{2}}}+4M_{N_{27,\frac{3}{2}}}\;,~~3M_{\Xi_{27,\frac{1}{2}}}+6M_{S_{27,1}}=8M_{N_{27,\frac{1}{2}}}+M_{N_{27,\frac{3}{2}}}\\
\label{3.10}&&3M_{\Omega_{27,1}}+9M_{S_{27,1}}=10M_{N_{27,\frac{1}{2}}}+2M_{N_{27,\frac{3}{2}}}
\end{eqnarray}
Both the masses of $\overline{\bf{35}}$-plets and the masses of
$\bf{27}$-plets contain three parameters:
$M_{\overline{35}},\alpha_{\overline{35}},\beta_{\overline{35}}$ or
$M_{27},\alpha_{27},\beta_{27}$, but we only known experimentally
the mass of  $S^{0}(3115)$ and $S^{+}(3140)$. So we can not fix the
masses of other nonaquark states which are  predicted by us.

It mostly seems that $S^{0}(3115)$ and $S^{+}(3140)$ would belong to
the same isospin multiplet and the mass difference
 between them
are mainly due to electromagnetic interaction and the mass
differences between u quark's and d quark's . \vskip0.3in
\begin{center}
{\subsection{The mass spectrum of the Nonaquark
$\overline{\bf{35}}$-plet}}
\end{center}
By means of the Gell-Mann--Okubo mass formula eq.(\ref{3.1}), and
noting the $\overline{\bf{35}}$-plet with $p=1,q=4$ whose weight
diagram is shown in Fig7., the masses of all the exotic nonaquark
states are as follows.
\begin{eqnarray}
\nonumber &&
M_{X_{\overline{35},\frac{1}{2}}}=M_{\overline{35}}+3\alpha_{\overline{35}}-\frac{11}{2}\beta_{\overline{35}}\;,
~~ M_{S_{\overline{35},1}}=M_{\overline{35}}+2\alpha_{\overline{35}}-3\beta_{\overline{35}}\;,~~M_{S_{\overline{35}}}=M_{\overline{35}}+2\alpha_{\overline{35}}-5\beta_{\overline{35}}\\
\nonumber
&&M_{N_{\overline{35},\frac{3}{2}}}=M_{\overline{35}}+\alpha_{\overline{35}}-\frac{\beta_{\overline{35}}}{2}\;,~~M_{N_{\overline{35},\frac{1}{2}}}=M_{\overline{35}}+\alpha_{\overline{35}}
-\frac{7}{2}\beta_{\overline{35}}\;,~~M_{\Sigma_{\overline{35},2}}=M_0+2\beta_{\overline{35}}\\
\nonumber&&M_{\Sigma_{\overline{35},1}}=M_{\overline{35}}-2\beta_{\overline{35}}\;,~~M_{\Xi_{\overline{35},\frac{5}{2}}}=M_{\overline{35}}-\alpha_{\overline{35}}+\frac{9}{2}\beta_{\overline{35}}\;,~~M_{\Xi_{\overline{35},\frac{3}{2}}}=M_{\overline{35}}-\alpha_{\overline{35}}-\frac{1}{2}\beta_{\overline{35}}\\
\label{3.2}&&M_{\Omega_{\overline{35},2}}=M_{\overline{35}}-2\alpha_{\overline{35}}+\beta_{\overline{35}}
\end{eqnarray}
We can find that the equal space rule holds for two sectors of
nonaquark baryons:
($X_{\overline{35},\frac{1}{2}},S_{\overline{35},1},N_{\overline{35},\frac{3}{2}},\Sigma_{\overline{35},2},\Xi_{\overline{35},\frac{5}{2}}$)
and
($S_{\overline{35}},N_{\overline{35},\frac{1}{2}},\Sigma_{\overline{35},1},\Xi_{\overline{35},
\frac{3}{2}},\Omega_{\overline{35},2}$). They satisfy the mass
relations as follows
\begin{equation}
\label{3.3}M_{X_{\overline{35},\frac{1}{2}}}-M_{S_{\overline{35},1}}=M_{S_{\overline{35},1}}
-M_{N_{\overline{35},\frac{3}{2}}}=M_{N_{\overline{35},\frac{3}{2}}}
-M_{\Sigma_{\overline{35},2}}=M_{\Sigma_{\overline{35},2}}-M_{\Xi_{\overline{35},\frac{5}{2}}}=
\alpha_{\overline{35}}-\frac{5}{2}\beta_{\overline{35}}
\end{equation}
\begin{equation}
\label{3.4}M_{S_{\overline{35}}}-M_{N_{\overline{35},\frac{1}{2}}}=M_{N_{\overline{35},\frac{1}{2}}}
-M_{\Sigma_{\overline{35},1}}=M_{\Sigma_{\overline{35},1}}-M_{\Xi_{\overline{35},\frac{3}{2}}}
=M_{\Xi_{\overline{35},\frac{3}{2}}}-M_{\Omega_{\overline{35},2}}=\alpha_{\overline{35}}
-\frac{3}{2}\beta_{\overline{35}}
\end{equation}
And we can also derive the mass relation between these nonaquark
states
\begin{eqnarray}
\nonumber&&M_{X_{\overline{35},\frac{1}{2}}}+M_{N_{\overline{35},\frac{3}{2}}}=2M_{S_{\overline{35},1}}\;,~~3(M_{S_{\overline{35}}}-M_{S_{\overline{35},1}})=2(M_{N_{\overline{35},\frac{1}{2}}}-M_{N_{\overline{35},\frac{3}{2}}})\\
\nonumber&&M_{\Sigma_{\overline{35},2}}+M_{S_{\overline{35},1}}=2M_{N_{\overline{35},\frac{3}{2}}}\;,~~M_{\Sigma_{\overline{35},1}}+3M_{S_{\overline{35},1}}=2(M_{S_{\overline{35}}}+M_{N_{\overline{35},\frac{3}{2}}})\\
\nonumber&&M_{\Xi_{\overline{35},\frac{5}{2}}}+2M_{S_{\overline{35},1}}=3M_{N_{\overline{35},\frac{3}{2}}}\;,~~2M_{\Xi_{\overline{35},\frac{3}{2}}}+9M_{S_{\overline{35},1}}=5M_{S_{\overline{35}}}+6M_{N_{\overline{35},\frac{3}{2}}}\\
\label{3.5}&&M_{\Omega_{\overline{35},2}}+6M_{S_{\overline{35},1}}=3M_{S_{\overline{35}}}+4M_{N_{\overline{35},\frac{3}{2}}}
\end{eqnarray}

\section{Decay of The Nonaquark States}

We assume the nonaquark state decays arise by a "fall-apart"
mechanism \cite{closecon,Jm,carlson,bs} without need for gluon
exchange to trigger the decay. There are some discussions on this
mechanism  in the pentaquark spectrum studies. By this mechanism, a
diquark in the pentaquark must be so clever that its two quarks are
detached to two isolated quarks,  and one of them enters into the
adjacent diquarks to form a baryon separatively, and another
combines with the residual anti-quark to form a meson, and then a
pentaquark baryon decays into a usual baryon plus a meson. In this
mechanism, the dynamics from the color coupling contributes  a
common factor \textcolor[rgb]{0.98,0.00,0.00}{to the decay amplitude
for a certain flavor multiplet } which is irrelevant to the
discussions on its decay branch fractions. Extending this
"fall-apart" mechanism to nonaquark state decays is natural and
straightforward: 1, one diquark is detached into two quarks; 2,
these two quarks enter the adjacent diquaks, and form two baryons
separatively; 3, the ninth quark also enter a diquark to form a
baryon, and then the nonaquark state consequently decays into three
baryons. We show this mechanism in the Fig5.
\begin{figure}[hptb]
\begin{center}
\includegraphics*[70pt,320pt][560pt,490pt]{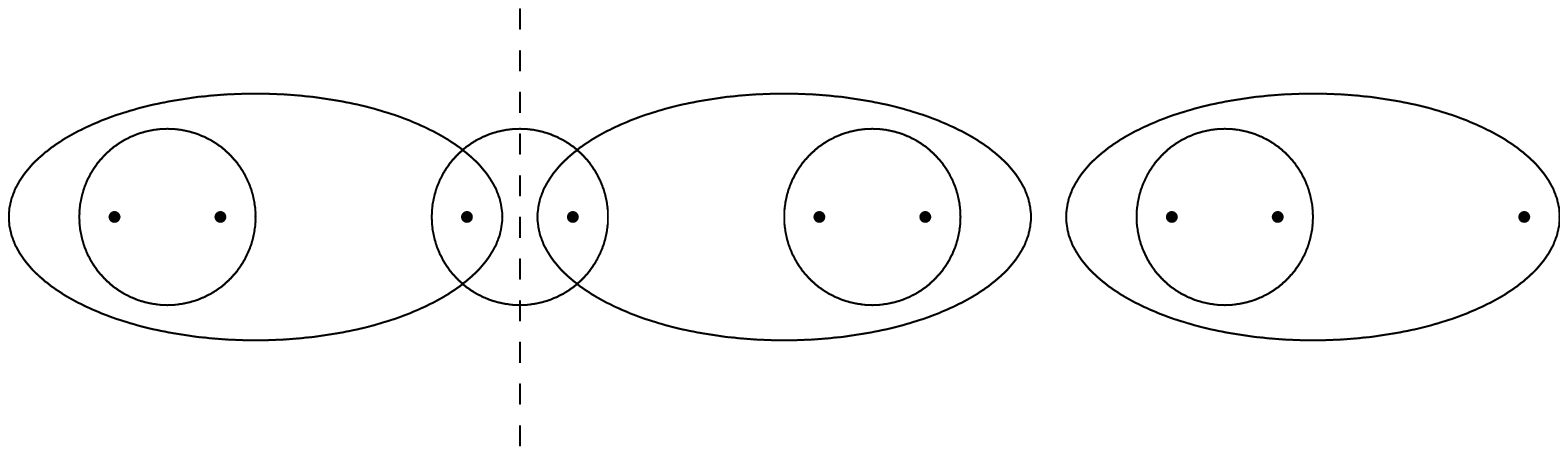}
\caption{The decay mechanism of Nonaquark states .}
\end{center}
\end{figure}

To the baryon octet, its quark content and the corresponding tensor
are well-known, which is listed in the Table I. The tensor basis of
the baryon octet is
\begin{equation}
\label{4.1}B^{l}_{r}=\frac{1}{\sqrt{3}}\varepsilon^{lmn}S_{rn}q_m
\end{equation}
with $S_{rn}=\frac{1}{\sqrt{2}}(q_rq_n+q_nq_r)$.
\begin{table}[hptb]
\caption{Baryon octet and its' flavor wave function }
\begin{center}
\begin{tabular}{|c|c|c|}\hline\hline
State & Tensor & Quark content\\\hline
p&$B^{3}_{1}$&$\frac{1}{\sqrt{2}}$[u,d]u\\\hline
n&$B^{3}_{2}$&$\frac{1}{\sqrt{2}}$[u,d]d\\\hline
$\Sigma^{+}$&$B^{2}_{1}$&$\frac{1}{\sqrt{2}}$[s,u]u\\\hline
$\Sigma^{0}$&$\frac{1}{\sqrt{2}}(B^{1}_{1}-B^{2}_{2})$&$\frac{1}{2}$([d,s]u+[u,s]d])\\\hline
$\Sigma^{-}$&$B^{1}_{2}$&$\frac{1}{\sqrt{2}}$[d,s]d\\\hline
$\Lambda^{0}$&-$\frac{3}{\sqrt{6}}B^{3}_{3}$&$\frac{1}{\sqrt{12}}$(2[u,d]s-[d,s]u-[s,u]d)\\\hline
$\Xi^{-}$&$B^{1}_{3}$&$\frac{1}{\sqrt{2}}$[d,s]s\\\hline
$\Xi^{0}$&$B^{2}_{3}$&$\frac{1}{\sqrt{2}}$[s,u]s\\\hline\hline
\end{tabular}
\end{center}
\end{table}
And the baryon singlet $\Lambda^{0}_{1}$ is given by
\begin{eqnarray}
\nonumber\Lambda^{0}_{1}&=&\frac{1}{2}\frac{1}{\sqrt{6}}\varepsilon^{klm}(q_lq_m-q_mq_l)q_k\\
\label{4.2}&=&\frac{1}{\sqrt{6}}([d,s]u+[s,u]d+[ u,d]s)
\end{eqnarray}
From these wave function we can see:
\begin{equation}
\label{4.3} \left\{
 \begin{array}{l}
  $[d,s]$u=\Sigma^{0}+\frac{\sqrt{6}}{3}\Lambda^{0}_{1}-\frac{\sqrt{3}}{3}\Lambda^{0}_{8}\\
  $[s,u]$d=-\Sigma^{0}+\frac{\sqrt{6}}{3}\Lambda^{0}_{1}-\frac{\sqrt{3}}{3}\Lambda^{0}_{8}\\
  $[u,d]$s=\frac{\sqrt{6}}{3}\Lambda^{0}_{1}+\frac{2\sqrt{3}}{3}\Lambda^{0}_{8}
 \end{array}
 \right.
\end{equation}

\begin{center}
{\subsection{The decay of $\bf{27}$-plet}}
\end{center}
Since the $\bf{27}$-plet is degenerate, i.e., $\bf{27}_1,~\bf{27}_2$
and $\bf{27}_3$ are the three irreducible representatives of
$\bf{27}-$dimension,  we discuss the decays for each case following
the method of dealing with the "fall-apart" decay \cite{close}.
Starting with the wave function of $S^{+}_{27_1,1}$ ( or
$S^{0}_{27_1,1}$
 etc.)  which is given in Eq.(\ref{2.33})(Eq.(\ref{2.34}) etc.), we rewrite the wave function in
the form of (qqq)(qqq)(qqq) which is similar to Ref.\cite{close},
then using Eq.(\ref{4.3}) and Table I and considering fermion
statistics, we can map $S^{+}_{27_1,1}$ ( or $S^{0}_{27_1,1}$ etc.)
into the ground state of three baryons, from which we can learn what
{particles are the  decay products of the $S^{+}_{27_1,1}$ (or
$S^{0}_{27_1,1}$ etc.), and derive further the ratios of the branch
fractions of these channels. We discuss the decays for each case
respectively by means of the above spirit of dealing with the decays
in the follows}:

\begin{enumerate}

\item The first case: The wave function of $S^{+}_{27_1,1}$ and
$S^{0}_{27_1,1}$ is given by Eq.(\ref{2.33}) and Eq.(\ref{2.34})
separately , and they are mapped into
\begin{equation}
\label{4.11} S^{+}_{27_1,1}\rightarrow\frac{1}{\sqrt{2}}\Sigma^{0}pn
\end{equation}
\begin{equation}
\label{4.12}S^{0}_{27_1,1}\rightarrow
\frac{1}{2}(\sqrt{2}\Sigma^{-}pn+\frac{\sqrt{6}}{3}\Lambda^{0}_{1}nn+\frac{2\sqrt{3}}{3}\Lambda^{0}_{8}nn).
\end{equation}
Then, we obtain that the main decay channel of $S^{+}_{27_1,1}$ is
$\Sigma NN$, and the ratio of coupling constants is
\begin{equation}
\label{4.13}g(S^{+}_{27_1,1}\Sigma^{-}pn):g(S^{+}_{27_1,1}\Lambda^{0}_{1}
nn):g(S^{+}_{27_1,1}\Lambda^{0}_{8}nn)
=\sqrt{2}:\frac{\sqrt{6}}{3}:\frac{2\sqrt{3}}{3}.
\end{equation}
Considering phase space effect, we get
\begin{equation}
\label{4.14}\frac{BR[S^{+}_{27_1,1}\rightarrow\Sigma
NN]}{BR[S^{+}_{27_1,1}\rightarrow\Lambda
NN]}=\frac{2}{((\frac{\sqrt{6}}{3})^2+\frac{2\sqrt{3}}{3})^2)\times4.164}\approx0.24
\end{equation}
Therefore,  $S^{0}_{27_1,1}$ dominantly decays into $\Lambda NN$,
and $S^{0}_{27_1,1}$  can not be $S^{0}(3115)$.

\item The second case: Eq.(\ref{2.37}) and Eq.(\ref{2.38}) give
the wave function of $S^{+}_{27_2,1}$ and  $S^{0}_{27_2,1}$ in this
case, and the map is as following
\begin{equation}
\label{4.15}S^{+}_{27_2,1}\rightarrow\frac{1}{2\sqrt{2}}(\Sigma^{+}nn+\Sigma^{-}pp+\sqrt{2}\Sigma^{0}pn),
\end{equation}
\begin{equation}
\label{4.16}S^{0}_{27_2,1}\rightarrow\frac{1}{2\sqrt{2}}(-\Sigma^{0}nn+\sqrt{2}
\Sigma^{-}pn-\sqrt{6}\Lambda^{0}_{1}nn-\sqrt{3}\Lambda^{0}_{8}nn).
\end{equation}
We note that in this case $S^{+}_{27_2,1}$ also can only decay into
$\Sigma NN$, and the ratio of the branch fractions
\begin{equation}
\label{4.17}BR[S^{+}_{27_2,1}\rightarrow\Sigma^{+}nn]:BR[S^{+}_{27,1}\rightarrow\Sigma^{-}pp]:BR[S^{+}_{27,1}\rightarrow\Sigma^{0}pn]=1:1:2
\end{equation}
and the main decay modes of $S^{0}_{27_2,1}$ is $\Lambda NN$
\begin{equation}
\label{4.18}\frac{BR[S^{0}_{27_2,1}\rightarrow\Sigma
NN]}{BR[S^{0}_{27_2,1}\rightarrow\Lambda NN]}\approx0.08
\end{equation}

\item The third case: The wave function of $S^{+}_{27_3,1}$ and
$S^{0}_{27_3,1}$ is given by Eq.(\ref{2.41}) and Eq.(\ref{2.42}),
similarly
\begin{equation}
\label{4.19}S^{+}_{27_3,1}\rightarrow\frac{3}{2}\Sigma^{0}pn+\frac{1}{2\sqrt{2}}
\Sigma^{+}nn+\frac{1}{2\sqrt{2}}\Sigma^{-}pp
\end{equation}
\begin{equation}
\label{4.20}S^{0}_{27_3,1}\rightarrow\frac{1}{4\sqrt{2}}(6\sqrt{2}\Sigma^{-}pn
-2\Sigma^{0}nn-\frac{2\sqrt{6}}{3}\Lambda^{0}_{1}nn+\frac{2\sqrt{3}}{3}\Lambda^{0}_{8}nn).
\end{equation}
From the above two formula, we know that in this case
$S^{+}_{27_3,1}$ decay mainly into $\Sigma NN$ not to $\Lambda
NN$,and
\begin{equation}
\label{4.21}BR[S^{+}_{27_3,1}\rightarrow\Sigma^{0}pn]:BR[S^{+}_{27_3,1}\rightarrow\Sigma^{+}nn]:BR[S^{+}_{27_3,1}\rightarrow\Sigma^{-}pp]=18:1:1
\end{equation}
and the ratio of branch fractions
\begin{equation}
\label{4.22}\frac{BR[S^{0}_{27_3,1}\rightarrow\Sigma
NN]}{BR[S^{0}_{27_3,1}\rightarrow\Lambda nn]}\approx2.8
\end{equation}
We  note that in this case the main decay mode of $S^{0}_{27_3,1}$
is $\Sigma NN$ which is consistent with the decay of $S^{0}(3115)$.
\end{enumerate}

\begin{center}
{\subsection {The decay of $\overline{\bf{35}}$-plet}}
\end{center}
Starting with the wave function of $S^{+}_{\overline{35},1}$ which
is given in Eq.(\ref{2.27}), we discuss the decays by means of the
method used in the previous sub-section to deal with {the decays}.
We  map $S^{+}_{\overline{35},1}$ into the ground state of three
baryons
\begin{equation}
\label{4.4} S^{+}_{\overline{35},1}\rightarrow\frac{1}{2}(\Sigma^{-}
pp+\Sigma^{+}nn)
\end{equation}
So, $S^{+}_{\overline{35},1}$ can  only decay into $\Sigma^{-}NN$,
but can not decay into $\Lambda^{-}NN$ (N stands for nucleon,i.e.,
proton or neutron). And the ratio of branch fractions is
\begin{equation}
\label{4.5}\frac{BR[S^{+}_{\overline{35},1}\rightarrow \Sigma^{-}
pp]}{BR[S^{+}_{\overline{35},1}\rightarrow \Sigma^{+}
nn]}=\frac{1}{1}=1
\end{equation}
Similarly $S^{+}_{\overline{35}}$ whose wave function is defined in
Eq.(\ref{2.28}) is mapped onto
\begin{equation}
\label{4.6}
S^{+}_{\overline{35}}\rightarrow\frac{1}{\sqrt{6}}(\Sigma^{-}
pp-\Sigma^{+}nn)
\end{equation}
also  $S^{+}_{\overline{35}}$ decay only into $\Sigma^{-}NN$, can
not decay into $\Lambda^{-}NN$, and the ratio of branch fractions is
\begin{equation}
\label{4.7}\frac{BR[S^{+}_{\overline{35}}\rightarrow \Sigma^{-}
pp]}{BR[S^{+}_{\overline{35}}\rightarrow \Sigma^{+}
nn]}=\frac{1}{1}=1
\end{equation}
We also obtain that the $S^{+}_{\overline{35}}\Sigma^{-}pp$ and
$S^{+}_{\overline{35}}\Sigma^{+}nn$ interactions have different
phase, while the $S^{+}_{\overline{35},1}\Sigma^{-}pp$ and
$S^{+}_{\overline{35},1}\Sigma^{+}nn$ interactions have the same
phase (see Eqs.(\ref{4.6}) and (\ref{4.4})).

 In the same way,
starting from Eq.(\ref{2.26}) which is the wave function of
$S^{0}_{\overline{35},1}$, we have:
\begin{equation}
\label{4.8}S^{0}_{\overline{35},1}\rightarrow\frac{1}{2}(-\Sigma^{0}+\sqrt{3}\Lambda^{0}_{8})nn
\end{equation}
So the main decay channel of $S^{0}_{\overline{35},1}$ is $\Lambda
NN$, but not $\Sigma NN$, i.e., ratio of the effective couplings
reads
\begin{equation}
\label{4.9}\frac{g(S^{0}_{\overline{35},1}\Sigma^{0}nn)}{g(S^{0}_{\overline{35},1}\Lambda^{0}_{8}nn)}=-\frac{1}{\sqrt{3}}.
\end{equation}
Considering the different three body phase space, we can obtain the
the ratio of branch fractions
\begin{equation}
\label{4.10}\frac{BR[S^{0}_{\overline{35},1}\rightarrow\Sigma^{0}nn]}{BR[S^{0}_{\overline{35},1}
\rightarrow\Lambda^{0}_{8}nn]}\approx0.05.
\end{equation}
So the observed  $S^{0}(3115)$ \cite{kek1} can not be $
S^{0}_{\overline{35},1}$. \vspace{0.5cm}

 From the decay of $\overline{\bf{35}}$-plet and
$\bf{27}$-plet, we see that the nonaquark state $S^{0}(3115)$ can
only possibly belong to $\bf{27}$-plet, and its flavor configuration
is Eq.(\ref{2.42}), or its main component is $S^{0}_{27_3,1}$ with
small mixing of $S^{0}_{27_1,1},S^{0}_{27_2,1},S^{0}_{35,1}$ . While
$S^{+}(3140)$ maybe belong to $\overline{\bf{35}}$-plet or
$\bf{27}$-plet, it possibly is
$S^{+}_{27_1,1},S^{+}_{27_2,1},S^{+}_{27_3,1},S^{+}_{35,1}$ or the
mixing of all of them, since these states have same quantum numbers.
From the discussion of subsection A in section II , we know that in
the exact $SU(3)^{flavor}\times SU(3)^{color}\times SU(2)^{spin}$
limit, $S^{0}(3115)$ and $S^{+}(3140)$ possibly belong to {\bf
27}-plet, this further give support to our suggestion that
$S^{0}(3115)$ can only possibly belong to $\bf{27}$-plet.
Furthermore in this case the nonaquark states have negative parity,
this is a unusual results, since "standard" nonaquark state which
involve 9 quarks in relative S-wave have positive parity, there
maybe really exists exotica nonaquark {\bf 27}-plet. We can obtain
useful information about $S^{+}(3140)$ by experimentally measuring
the branch fractions of its decay channels. \vskip0.3in
\begin{center}
{\section{Discussion and Conclusion}}
\end{center}
\vskip-0.5in In summary, we have obtain the wave functions of the
nonaquark states in $SU(3)$ quark model with diquark correlation
using standard direct tensor decomposition, we predict the existence
of other nonaquark states.  It would be helpful for constructing the
effective interaction Lagrangian to describe the nonaquark decays
with rational $SU(3)-$flavor structure. We obtain some interesting
mass sum rules for the nonaquark $\overline{\bf{35}}$-plet and
$\bf{27}$-plet, but it is still open to fix the spectrum of
$\overline{\bf{35}}$-plet and $\bf{27}$-plet due to the scarcity of
experiments. More data are expected. Under the assumption of the
"fall-apart" decay mechanism which has been subtly used in studying
pentaquark decays, we find out  that the $S^{0}(3115)$
 belongs to a $\bf{27}$-plet. Its main component is
$S^{0}_{27_3,1}$, and the mixing with
$S^{0}_{27_1,1},S^{0}_{27_2,1},S^{0}_{35,1}$ must be few. It is
possible that $S^{0}(3115)$ and $S^{+}(3140)$ belong to the same
isospin multiplet, since their mass difference is about 25 MeV which
can be interpreted  by electromagnetic interaction and u,d quark
mass difference. Further in the exact $SU(3)^{flavor} \times
SU(3)^{color} \times SU(2)^{spin}$ limit, $S^{0}(3115)$ and
$S^{+}(3140)$ are belong to the {\bf 27}-plet, and its parity is
negative. We suggest to study the decays of $S^{+}(3140)$ more in
experiment, especially the ratios of the various decay mode's branch
fractions, through which we can learn the structure of $S^{+}(3140)$
and maybe discover new nonaquark state .

There maybe other types of quark correlation in nonaquark state,
such as (qqq)-(qqq)-(qqq), since either (qq$\bar{\rm{q}}$)-(qq)
quark correlation \cite{lipkin} or diquark correlation \cite{jaffe}
is possible in pentaquark, they have different flavor and color
structures from the diquark correlation, {so the masses, the decay
property, and the products etc of the states may be different} from
the prediction of quark model with diquark correlation.

It would be interesting to study the mixing between the nonaquark
states with same quantum numbers. It is necessary to considering the
mixing in order to  compare theoretical prediction with experimental
data more precisely. This would be highly nontrivial.

Finally, we might image that this multi-baryon problem could be
studied in other quark models, e.g., the model in \cite{swart}, and
in the chiral soliton model\cite{skyrme}.

\begin{center}
{\bf ACKNOWLEDGMENTS}
\end{center}
We acknowledge Dr. J. Deng for his help in plotting the diagrams. We
would like also to thank the referee for his suggestion to discuss
the color symmetry in the nonaquark wavefunction. This work is
partially supported by National Natural Science Foundation of China
under Grant Numbers 90403021, and by the PhD Program Funds of the
Education Ministry of China and KJCX2-SW-N10 of the Chinese Academy.

\begin{figure}[hptb]
\begin{center}
\includegraphics*[100pt,200pt][525pt,600pt]{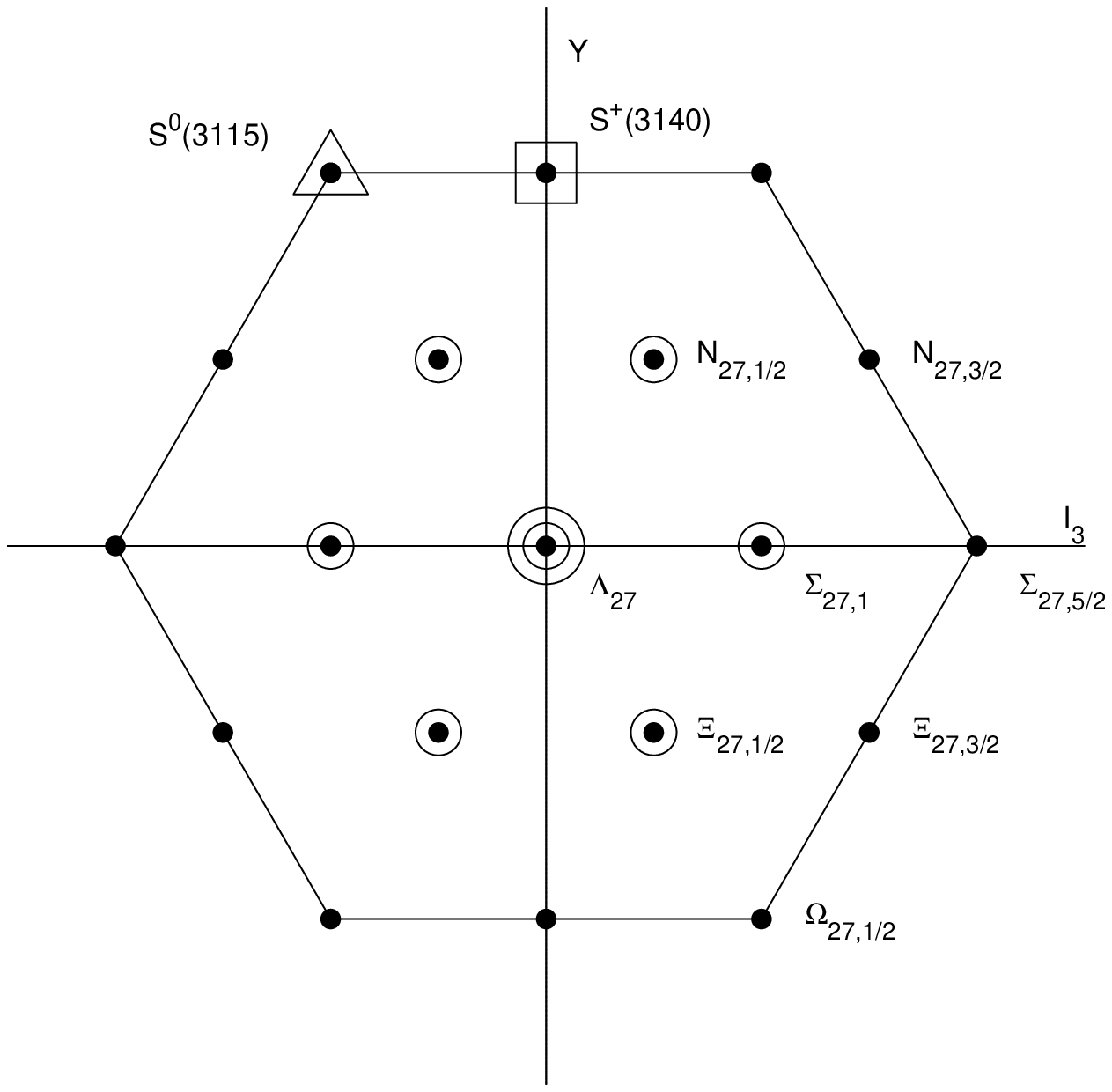}
\caption{The weight diagram of nonaquark $\bf{27}$-plet, and the
location of $S^{0}(3115)$ and $S^{+}(3140)$ are expressly shown by
the triangle and square respectivelly }.
\end{center}
\end{figure}
\begin{figure}[hptb]
\begin{center}
\includegraphics*[75pt,200pt][565pt,605pt]{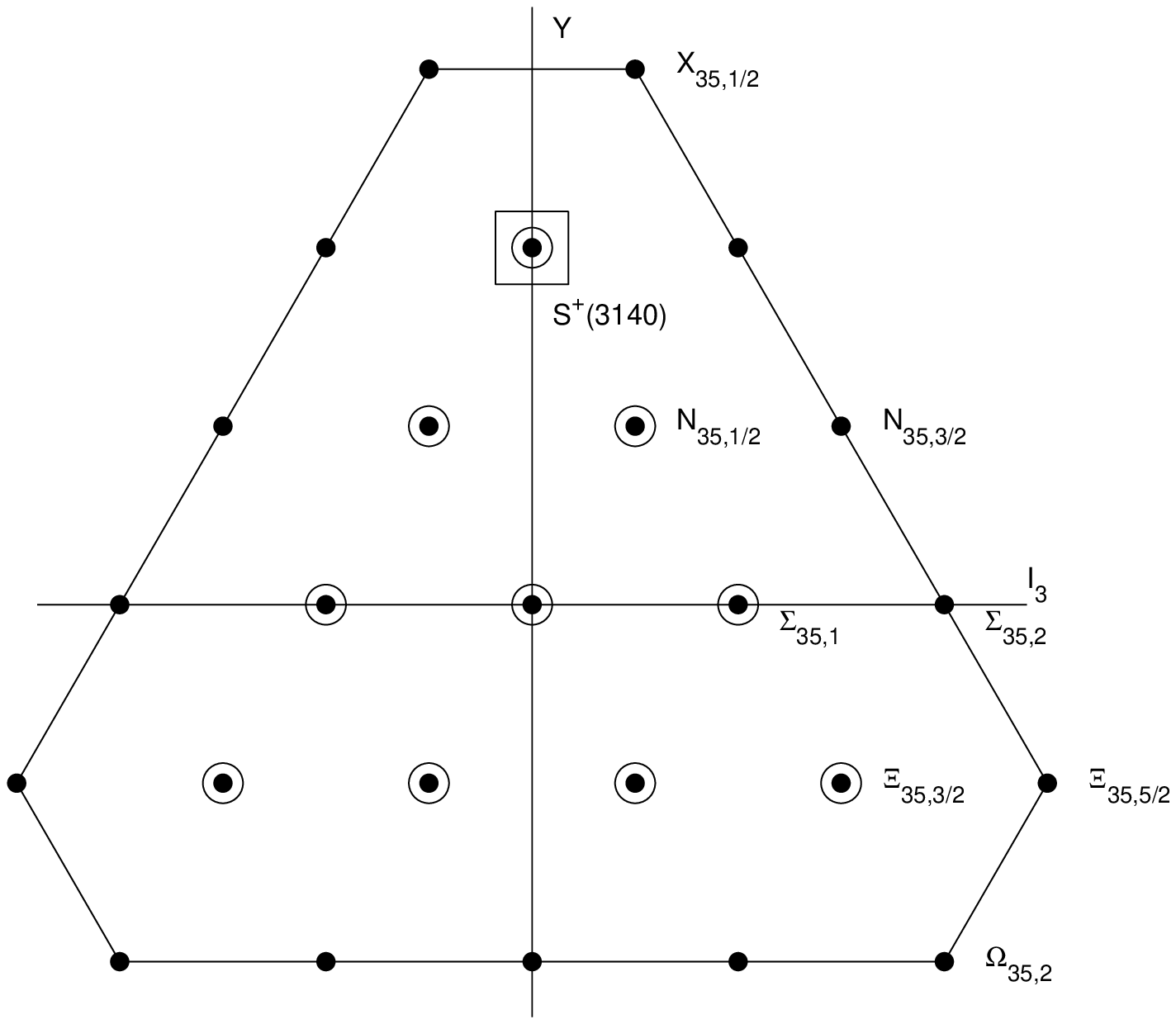}
\caption{The weight diagram of nonaquark $\overline{\bf{35}}$-plet,
only $S^{+}(3140)$ is specially shown by square, whereas
$S^{0}(3115)$ is not shown. Because our analysis indicate that
$S^{0}(3115)$ can not belong to $\overline{\bf{35}}$-plet }.
\end{center}
\end{figure}
\end{document}